\documentclass[aps,prx,twocolumn,superscriptaddress,longbibliography]{revtex4-2}
\usepackage{graphicx,amsmath,amssymb,booktabs,xcolor,hyperref}
\usepackage{tikz}
\usetikzlibrary{quantikz2}
\graphicspath{{./}}
\newcommand{\rev}[1]{#1}
\newcommand{\nkd}[3]{[\![#1,#2,#3]\!]}

\begin{document}
\title{Singlet flags: using native entanglement to restore the circuit distance of shuttled qLDPC
codes in silicon}
\author{William Chow}
\affiliation{Institute for Quantum Computing and Department of Physics, University of Waterloo, 200
University Avenue West, Waterloo, Ontario N2L 3G1, Canada}
\author{Jonathan Baugh}
\affiliation{Institute for Quantum Computing and Department of Chemistry, University of Waterloo,
200 University Avenue West, Waterloo, Ontario N2L 3G1, Canada}
\date{\today}

\begin{abstract}
High-rate quantum low-density parity-check (qLDPC) codes could greatly reduce the number of physical
qubits a silicon spin-qubit processor needs, but their checks are long-range and of high weight.
Electron shuttling offers two ways to measure such checks: carry one ancilla electron through all
the data qubits of a check, or distribute spin singlets between static nodes and measure the check
through entanglement.
In earlier work we compared the two on a common layout and found that each wins in a different
regime.
Here we combine them.
The mobile ancilla is made one half of a spin singlet whose partner stays at home, and a Bell
measurement at the end of the check \rev{provides a flag for isolated bit-flip faults acquired
during the walk.}
This targets the main weakness of the mobile ancilla: a single fault on it spreads to several data
qubits, so the code's \emph{circuit} distance can be far below its nominal distance.
We ask whether the flag only lowers the prefactor of the logical error rate or also raises its
exponent, which the circuit distance sets.
To measure this cleanly we construct distance-6 codes with check weights 8, 10 and 12 and compute
their circuit distances exactly with two independent solvers.
Without the flag, circuit distance falls as check weight grows, to as low as 1 at weight 12.
With it, every weight-8 and weight-10 code we tested recovers circuit distance 5 or 6, which corrects
as many faults as the code itself.
For codes of distance 8 to 12, the largest we study, CAbLECAR keeps only three-eighths to
three-quarters of the code distance, and the flag raises the circuit distance by one to four: the BB
$\nkd{108}{8}{10}$ code regains its full distance of 10, and the Gross code $\nkd{144}{12}{12}$ rises
from 7 to 10 in both bases.
In circuit-level simulations with a Pauli noise model of shuttled spin qubits, the flag lowers
logical error rates by typically $2$--$10\times$
across bivariate-bicycle, circulant-permutation and surface codes, including codes of distance 8 to
12, and the gain survives \rev{percent-level flag-readout errors and percent-level bit flips on the
electrons of the singlet.}
The exact distances also show that BP+OSD does not reach the full circuit distance: it fails on fault
pairs, and at weight 12 on single faults, that the flagged circuit can correct, so our flagged
logical error rates are upper bounds.
Decoding the same shots exactly raises the flag's gain on the distance-6 codes from $3$--$7\times$
to $5$--$17\times$ for the weight-6 BB code, and from $1$--$12\times$ to $5$--$90\times$ at weights 8
and 10 for $p\ge2.5\times10^{-4}$.
In particular, for the weight-10 code under uniform noise BP+OSD finds no gain in the $Z$ basis at low
error rates, but exact decoding finds $16$--$39\times$.
\end{abstract}
\maketitle

\section{Introduction}

Quantum error correction will dominate the qubit budget of any useful quantum computer.
Surface codes~\cite{kitaev2003} need hundreds to thousands of physical qubits per logical qubit.
Quantum low-density parity-check (qLDPC) codes can do much better.
The bivariate-bicycle (BB) codes~\cite{bravyi2024} brought this into view: the $\nkd{144}{12}{12}$
``Gross'' code, with checks of weight 6, protects 12 logical qubits with 288 physical qubits, where
the surface code would need nearly 3000.
Since then, codes of considerably higher rate have appeared.
Pair-partition constructions from circulant permutation matrices~\cite{okada2026,ultrahighrate2026}
reach encoding rates approaching $k/n=0.5$, and BB-type codes with weight-8 checks improve on the BB
parameters~\cite{weight8bb2026}.
These \emph{ultra-high-rate} codes buy their rate with checks heavier than the BB codes' weight 6,
typically 8 to 12.
For silicon spin qubits the rate matters even though the qubits themselves are tiny: every qubit
still needs gate electrodes, wiring, control electronics and chip area.

The catch is connectivity.
A qLDPC code with a good rate needs checks that involve 6 to 12 data qubits spread across the chip,
while spin qubits interact only with their neighbours.
Electron shuttling, moving electrons coherently along conveyor
lanes~\cite{seidler2022,langrock2023,desmet2025}, is the natural fix, and it can be used in two
quite different ways.

\emph{Mobile ancillas.}
An ancilla electron is shuttled to each data qubit of a check in turn and interacts with it, then
returns to be read out.
CAbLECAR~\cite{cablecar} compiles complete syndrome-extraction rounds of BB codes in this way.
Other schedules for qLDPC codes on shuttling spin qubits have since appeared, from routing studies of
the Gross code~\cite{incerti2026} to codes co-designed with nested shuttling tracks~\cite{siegel2026nest}.

\emph{Singlet fabric.}
In our earlier work~\cite{chow2026syndrome} the data never move and neither do ancillas.
Instead, spin singlets, which can already be separated by shuttling while staying
entangled~\cite{struck2024}, are created on the shuttling lanes and delivered to static nodes, where
local gates consume them to measure non-local checks, for instance by building a GHZ state across
the nodes of a check.

Comparing the two under identical hardware parameters~\cite{chow2026syndrome}, we found that each wins
in a different regime.
On a common layout with two-qubit gate and readout errors of $10^{-3}$, the mobile ancilla gives the
lower logical error rate at every sampled shuttle-noise level for the Gross code, and when shuttling
causes no bit flips it always wins.
The fabric wins when shuttling frequently flips spins (probability $p_{\rm sh}/3$ per hop) and
two-qubit gate and readout errors are below about $4$--$5\times10^{-4}$; with rarer bit flips 
($p_{\rm sh}/10$) the two become comparable only near $10^{-4}$--$2\times10^{-4}$.
The fabric's advantage comes from where its transport happens: singlet halves are moved
\emph{before} they touch any data, so a transport fault stays on the singlet, whereas a travelling
ancilla copies a bit flip onto every data qubit it visits afterwards.

That observation is the starting point of this paper.
A transport fault on the mobile ancilla is dangerous because it propagates, and the classical fix
for propagating ancilla faults is a \emph{flag}~\cite{chao2018,chamberland2018}: an extra qubit that
detects when the ancilla has gone wrong.
Silicon devices that produce singlets already have a free source of flags.
We therefore ask whether we can keep the efficiency of the mobile ancilla and borrow the
entanglement of the singlet fabric to catch its faults.

Our proposal, the \emph{singlet flag}, does exactly this.
The mobile ancilla is created as one half of a spin singlet.
Its partner stays at the ancilla's home node while the ancilla walks its check.
When the ancilla returns, a Bell measurement on the pair reveals whether the ancilla's spin flipped
along the way (Sec.~\ref{sec:scheme}).

The figure of merit we care most about is the \emph{circuit distance}: the smallest number of faults
anywhere in the syndrome-extraction circuit that cause an undetected logical error.
A single fault on a mobile ancilla can corrupt many data qubits at once, so the circuit distance can
be much lower than the code distance $d$, and this gap grows with check weight
(Sec.~\ref{sec:distance-intro}).

A flag can lower the logical error rate in two different ways.
Below threshold, $P_L\approx A\,(p/p_{\rm th})^{\lceil d_{\rm circ}/2\rceil}$.
Catching likely faults lowers the prefactor $A$, a constant factor at every $p$; restoring circuit
distance raises the exponent, a gain that grows as $p$ falls.
We want to know which of the two the singlet flag achieves.
Exact circuit distances answer this independently of any decoder, and circuit-level simulations
measure the logical error rate itself.
All our results assume circuit-level Pauli noise: depolarising noise after every gate, bit-flip
errors on preparation, readout and singlet generation, phase and bit flips on every shuttle hop, and
dephasing of idle qubits (Sec.~\ref{sec:noise}).

Our main results are:
\begin{enumerate}
\item \textbf{Circuit distance.}
Using new distance-6 codes with check weights 8, 10 and 12 (Sec.~\ref{sec:codes}) and two
independent exact solvers (Sec.~\ref{sec:distance}), we show that without the flag the circuit
distance falls with check weight, reaching 1 at weight 12.
With the flag, every weight-8 and weight-10 instance reaches circuit distance 5 or 6, so it corrects
two faults, as many as the code.
For codes of distance 8 to 12, the largest we study, CAbLECAR keeps only three-eighths to
three-quarters of the code distance.
The flag raises the circuit distance by one to four (Table~\ref{tab:distlarge}): it lifts the
$\nkd{176}{48}{10}$ code from 5 and 6 to 7, the Gross code from 7 to 10 and $\nkd{180}{42}{12}$ from 8
to 10 in both bases, and restores the full distance of 10 to the BB $\nkd{108}{8}{10}$ code, from 6
and 7.
\item \textbf{Logical error rates.}
When shuttling flips spins, the flag lowers the logical error rate by $2$--$8\times$ for BB codes and
by up to $\sim\!10\times$ for surface codes (Sec.~\ref{sec:mc}), and the
gain survives \rev{flag-readout errors and singlet-preparation bit flips of a few percent
(Sec.~\ref{sec:robust}).}
With every error rate scaled together, the weight-8 code gains $8$--$24\times$ in the $X$ basis and
$2.8$--$5.6\times$ in the $Z$ basis for $p\le10^{-3}$, and codes of distance 8, 10 and 12 gain
$1.6$--$8.4\times$, with the gain growing as $p$ falls at distances 8 and 10 over the range we
sample.
With BP+OSD the weight-10 code gains about $5\times$ in the $X$ basis but only matches CAbLECAR in
the $Z$ basis below $p\approx3\times10^{-4}$, a limit of the decoder (item 3).
Without transport bit flips the Gross code loses at high transport phase-flip rates.
\item \textbf{Decoder gap.}
Using the exact circuit distances, we show that BP+OSD does not reach the full circuit distance the
flag restores (Sec.~\ref{sec:decoder}), as others have found for other codes~\cite{fregona2026,benito2025}.
At weight 8 it fails on 22 fault pairs that the flagged circuit can correct, which give about 40\%
of the flag's logical error rate at the lowest error rates we simulate, and at weight 12 it fails on
single faults.
Higher OSD order and BP+LSD do not help, but exact minimum-weight decoding removes every such failure.
Decoding the same sampled shots exactly, the flag's gain over CAbLECAR rises from $3$--$7\times$ to
$5$--$17\times$ for BB72, from $2$--$12\times$ to $5$--$70\times$ at weight 8, and from $1$--$5\times$
to $13$--$90\times$ at weight 10 for $p\ge2.5\times10^{-4}$.
In the $X$ basis this gain grows as $p$ falls, at weight 10 from $13\times$ at $p=10^{-3}$ to about
$270\times$ at $1.25\times10^{-4}$, while with BP+OSD it stays near $4$--$5\times$.
\end{enumerate}

\section{Why mobile ancillas lose distance}\label{sec:distance-intro}

Consider an $X$-type check of weight $w$, measured by an ancilla that couples to data qubits
$q_1,\dots,q_w$ in that order, acting as the control of each two-qubit gate.
If the ancilla suffers an $X$ error after the $j$-th gate, the error is copied onto the $w-j$ data
qubits it has yet to visit.
Up to multiplication by the check itself, this is a data error of weight $\min(j,\,w-j)$, which can
be as large as $\lfloor w/2\rfloor$.
Such correlated errors are called \emph{hook} errors.

For the weight-4 checks of the surface code, hooks have weight at most 2, and choosing the gate
order carefully makes them harmless~\cite{tomita2014}.
For qLDPC checks of weight 8 to 12, a single hook can have weight 4 to 6, comparable to the code
distance itself\rev{; in the schedules studied here, these faults can reduce the circuit
distance.}

The standard remedy for hook errors is a \emph{flag}.
Chao and Reichardt showed that a single extra qubit, coupled to the ancilla at the start and end of
a check, detects any fault that would spread into a high-weight data error, so that several
distance-3 codes, including the $[[5,1,3]]$ and $[[7,1,3]]$ codes, can correct any single fault with
only two extra qubits~\cite{chao2018}.
Chamberland and Beverland extended flag error correction to codes of arbitrary
distance~\cite{chamberland2018}, and Chao and Reichardt later gave a flag scheme for any stabilizer
code~\cite{chao2020}.
Flags have since become an architectural tool.
IBM's heavy-hexagon and heavy-square layouts place flag qubits on the edges of a low-degree lattice
and feed their outcomes to the decoder, so that the codes reach their full distance despite limited
connectivity~\cite{chamberland2020}.
Flag and bridge qubits have been used to map codes onto near-term processors~\cite{lao2020}, and
flag-proxy networks apply the same idea to qLDPC codes on hardware of bounded
degree~\cite{vittal2024}.
Flag gadgets built from classical codes cut the number of extra qubits needed at arbitrary
distance~\cite{anker2022}, and syndrome-extraction circuits with a few ancillas per check are provably
fault tolerant for weight-6 qLDPC checks~\cite{khesin2026}.
All of these flags are static qubits wired to the ancilla by extra two-qubit gates during the check.
The singlet flag differs in two ways: its entanglement with the ancilla is created \emph{before} the
check, by the hardware's native singlet source, and it never interacts with the ancilla again until
the ancilla returns home.
It therefore adds no gates to the walk itself.
Using the second spin of a pair is not new in silicon: readout already relies on a reference
spin~\cite{hetenyi2024}, and Siegel and Benjamin use the two-spin structure of singlet-triplet qubits
to turn leakage into erasures in shuttling-based surface codes~\cite{siegel2026erasure}.
To our knowledge, the partner spin has not been used as a flag.

Two recent proposals reduce hook errors in qLDPC circuits without flags.
NeST~\cite{siegel2026nest} chooses gate orders, together with enough freedom in the shuttling schedule,
so that the circuit distance comes close to the code distance.
Bi, Chang and Puri~\cite{bi2026} show that when ancillas suffer only phase flips, the standard BB
circuit has a markedly higher fault distance, and report close to an order of magnitude lower logical
error rate at $2\times10^{-3}$ for an ancilla bias of 50.
Both are complementary to the flag.
Gate ordering constrains the schedule, and bias helps only to the extent that transport is free of
bit flips, whereas the singlet flag works with any gate order and targets exactly the transport bit
flips that turn a travelling ancilla's faults into hooks.

We quantify the damage by the circuit distance
\begin{equation}
d_{\rm circ}=\min\{\,|e| \;:\; He=0,\ Le\neq0\,\},\label{eq:dcirc}
\end{equation}
where $e$ is a set of fault mechanisms from the circuit's detector error model, $H$ records which
detectors each mechanism fires and $L$ which logical observables it flips.
In words, $d_{\rm circ}$ is the fewest faults that flip a logical while firing no detector.
\rev{Any set of at most $t=\lfloor (d_{\rm circ}-1)/2\rfloor$ faults is correctable, in that no other set of
at most $t$ faults has the same syndrome and a different logical effect, so} for a distance-6 code a
circuit distance of 5 or 6 keeps the full $t=2$, while 3 or 4 drops it to $t=1$.

\section{The singlet flag}\label{sec:scheme}

\begin{figure}[t]\centering
\includegraphics[width=\columnwidth]{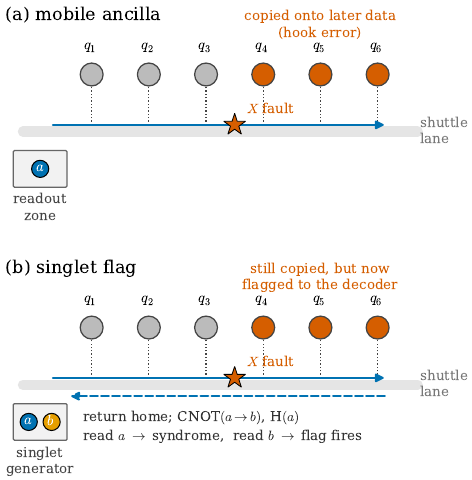}
\caption{Mobile-ancilla syndrome extraction.
(a) A bit flip on the travelling ancilla $a$ partway through its check is copied onto every data
qubit it visits afterwards, \rev{producing a hook error without a dedicated ancilla-fault flag.}
(b) With a singlet flag, $a$ starts as one half of a singlet whose partner $b$ waits in the
generator at the home node.
The hook still occurs, but on return the Bell measurement of the pair reveals it: $a$ gives the
syndrome bit and $b$ the flag.}\label{fig:scheme}
\end{figure}

\subsection{Protocol}

\begin{figure}[t]\centering
\resizebox{\columnwidth}{!}{%
\begin{quantikz}[row sep={0.66cm,between origins}, column sep=0.3cm]
\lstick{$a$} & \gate{H}\gategroup[2,steps=2,style={dashed,rounded corners,fill=blue!7,inner xsep=2pt},background,label style={label position=above,anchor=south,yshift=-0.05cm}]{\footnotesize singlet} & \ctrl{1} & & \ctrl{2}\gategroup[5,steps=5,style={dashed,rounded corners,fill=orange!9,inner xsep=2pt},background,label style={label position=above,anchor=south,yshift=-0.05cm}]{\footnotesize walk ($X$ check)} & \ctrl{3} & \gate[style={fill=red!20,draw=red!70!black}]{X} & \push{\ \cdots\ } & \ctrl{4} & & \ctrl{1}\gategroup[2,steps=3,style={dashed,rounded corners,fill=green!9,inner xsep=2pt},background,label style={label position=above,anchor=south,yshift=-0.05cm}]{\footnotesize Bell measurement} & \gate{H} & \meter{} & \rstick{syndrome}\\
\lstick{$b$} & & \targ{} & & & & & & & & \targ{} & & \meter{} & \rstick{flag}\\
\lstick{$q_1$} & & & & \targ{} & & & & & & & & & \\
\lstick{$q_2$} & & & & & \targ{} & & & & & & & & \\
\lstick{$q_w$} & & & & & & & & \targ{} & & & & &
\end{quantikz}}
\caption{Circuit of one singlet-flagged $X$ check of weight $w$.
The generator prepares $a$ and $b$ in the Bell state $|\Phi^+\rangle$; $a$ then applies a CNOT to
each data qubit $q_i$ in its visit order, and on return a Bell measurement gives the syndrome ($a$)
and the flag ($b$).
Each CNOT is a CZ with Hadamards on its target; a $Z$ check omits the Hadamards on the data.
The red $X$ marks a bit flip on $a$ during transport, after the second gate: it is copied onto
$q_3,\dots,q_w$ and, as derived in Sec.~\ref{sec:heisenberg}, it also flips the flag.}\label{fig:circuit}
\end{figure}
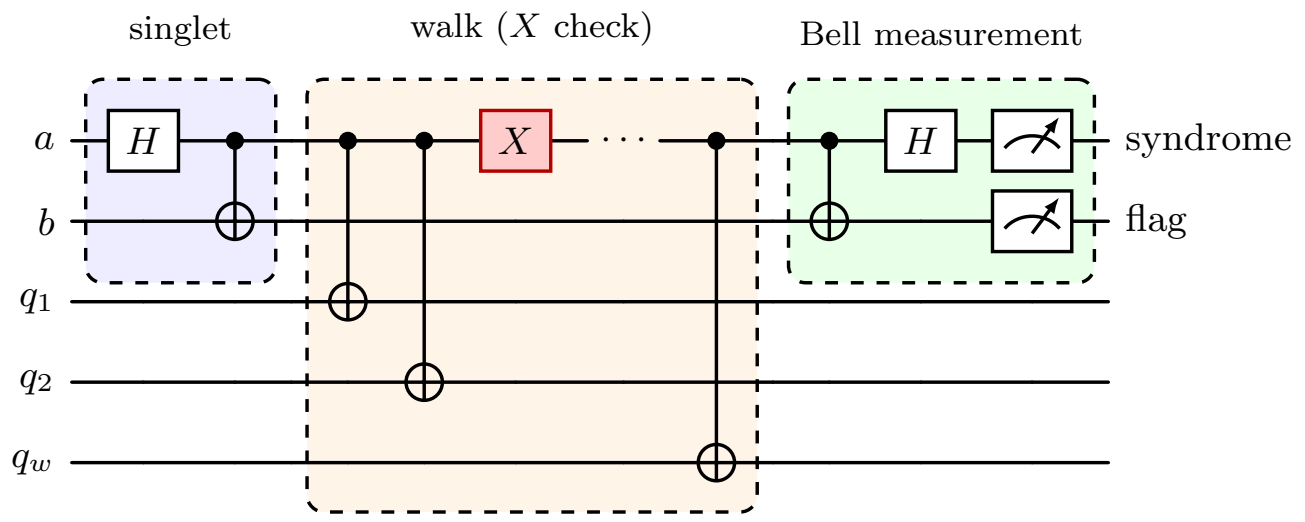

We build on the CAbLECAR architecture~\cite{cablecar}: a grid of static nodes connected by shuttling
lanes, with interaction zones for two-qubit gates and readout zones with charge sensors.
Each check has one ancilla, which is routed through the data qubits of its check and back.
We add one component per node: a \emph{singlet generator}, adjacent to the readout zone and with its
own charge sensor.
Each check then runs in three stages (Figs.~\ref{fig:scheme} and~\ref{fig:circuit}).
\begin{enumerate}
\item \emph{Pair creation.} The generator prepares two electrons $a$ and $b$ in a Bell state.
Electron $a$ is released to act as the ancilla; $b$ stays in the generator.
\item \emph{Walk.} Electron $a$ visits the data qubits of its check in turn, exactly like a
CAbLECAR ancilla, and applies a CZ, our native two-qubit gate, with each, as the control.
For an $X$ check, Hadamards on the data qubit before and after the CZ turn it into a CNOT.
\item \emph{Bell measurement.} Electron $a$ returns home.
A CNOT from $a$ to $b$ (a CZ with Hadamards on $b$) and a Hadamard on $a$ rotate the pair into the
measurement basis.
Reading $a$ gives the syndrome bit and reading $b$ gives the flag bit.
\end{enumerate}
\rev{Both readouts are single-spin measurements in the $Z$ basis by Pauli spin blockade against a static
reference spin~\cite{hetenyi2024}, so every sensor has its own reference electron: the readout zone
of each node that reads an ancilla, as in CAbLECAR, and the generator, which reads $b$.
The two readouts of a flagged check start together after the Bell rotation and run in parallel on
the two sensors, so the flag adds no readout time to the round.
We treat each readout as an effective operation of 500\,ns, CAbLECAR's value, that includes
spin-to-charge conversion, charge sensing and the reset of the reference spin, and that reports the
wrong outcome with probability $p$ (Sec.~\ref{sec:noise}).}

\subsection{Why the flag works}\label{sec:heisenberg}

The clearest way to see what the two readouts measure is to follow the relevant Pauli operators
through the circuit, in the Heisenberg picture.
The generator prepares $|\Phi^+\rangle_{ab}=(|00\rangle+|11\rangle)/\sqrt2$, which is the $+1$
eigenstate of
\begin{equation}
X_aX_b \quad\text{and}\quad Z_aZ_b .
\end{equation}
(The native singlet differs from $|\Phi^+\rangle$ by a fixed Pauli on $b$, which only relabels the
expected outcomes.)
Let $S=P_{q_1}P_{q_2}\cdots P_{q_w}$ be the check being measured, with $P=X$ for an $X$ check and
$P=Z$ for a $Z$ check.

\paragraph*{The walk.} Each walk gate has $a$ as its control, so it acts on the ancilla's Paulis as
\begin{equation}
X_a \;\mapsto\; X_a\,P_{q_i}, \qquad Z_a \;\mapsto\; Z_a ,\label{eq:walkgate}
\end{equation}
for $X$ checks ($P=X$) and $Z$ checks ($P=Z$) alike.
Applying Eq.~\eqref{eq:walkgate} for every data qubit of the check gives
\begin{equation}
X_aX_b \;\mapsto\; X_aX_b\,S, \qquad Z_aZ_b \;\mapsto\; Z_aZ_b .\label{eq:walk}
\end{equation}
The first relation is what makes the circuit measure the check: after the walk, the pair's $XX$
parity equals the eigenvalue $s=\pm1$ of $S$ on the data.
The second relation is what makes the flag possible: the pair's $ZZ$ parity is untouched by the
walk and is still $+1$.

\paragraph*{The Bell measurement.} At the end, $a$ and $b$ are read in the $Z$ basis after
$\mathrm{CNOT}_{a\to b}$ and $H_a$.
Tracing the two measured operators back through these gates,
\begin{equation}
\begin{aligned}
Z_a &\;\xleftarrow{\;H_a\;}\; X_a \;\xleftarrow{\;\mathrm{CNOT}_{a\to b}\;}\; X_aX_b,\\
Z_b &\;\xleftarrow{\;H_a\;}\; Z_b \;\xleftarrow{\;\mathrm{CNOT}_{a\to b}\;}\; Z_aZ_b .
\end{aligned}\label{eq:bell}
\end{equation}
So reading $a$ measures $X_aX_b$, which by Eq.~\eqref{eq:walk} is the syndrome $s$, and reading $b$
measures $Z_aZ_b$, which is $+1$ without faults.
The flag is therefore a deterministic detector: it should never fire.

\paragraph*{Faults.} A Pauli fault on $a$ or $b$ before the Bell measurement is pushed forward by
the remaining gates and then flips whichever readout it anticommutes with
(Table~\ref{tab:faults}).
\begin{itemize}
\item A \emph{bit flip} $X_a$ after the $j$-th gate is copied by Eq.~\eqref{eq:walkgate} onto the
remaining data, $X_a\mapsto X_a P_{q_{j+1}}\cdots P_{q_w}$: this is the hook.
But $X_a$ anticommutes with $Z_aZ_b$, so the flag fires.
The hook is still there; it is no longer silent.
\item A \emph{phase flip} $Z_a$ commutes with every walk gate, so it never reaches the data.
It anticommutes with $X_aX_b$ and flips the syndrome bit, exactly like a measurement error on an
ordinary ancilla, which repeated rounds already handle.
\item Faults on the static partner act only on the readouts: $X_b$ raises a false flag and $Z_b$
flips the syndrome bit.
\end{itemize}
The flag counts bit flips on $a$ only modulo 2.
Any odd number of them fires it, but any even number goes unnoticed: two flips, after gates $j$ and
$k>j$, cancel in the flag ($X_a^2=I$) yet leave the data error $P_{q_{j+1}}\cdots P_{q_k}$ on the
stretch of the walk between them, and $2m$ flips leave data errors on $m$ such stretches.

\begin{table}[t]
\caption{Effect of faults on the pair during one flagged check, from
Eqs.~\eqref{eq:walkgate}--\eqref{eq:bell}.
The flag fires on any odd number of bit flips on the travelling ancilla $a$, including the single
hook of the first row.
An even number of flips cancels in the flag but still leaves data errors on the stretches of the walk
between them; the last row shows two.
$P=X$ for $X$ checks and $P=Z$ for $Z$ checks.}\label{tab:faults}
\begin{ruledtabular}\begin{tabular}{lccc}
Fault & Syndrome & Flag & Data error\\\hline
$X_a$ after gate $j$ & -- & flips & $P_{q_{j+1}}\cdots P_{q_w}$\\
$Z_a$ & flips & -- & --\\
$X_b$ & -- & flips & --\\
$Z_b$ & flips & -- & --\\
$X_a$ after gates $j<k$ & -- & -- & $P_{q_{j+1}}\cdots P_{q_k}$\\
\end{tabular}\end{ruledtabular}
\end{table}

\subsection{Cost}

The flag needs no extra shuttled electrons and no extra lanes: \rev{one generator per node, with its own
sensor and reference spin}, and per check one extra CZ, three Hadamards and one readout.
Because the ancilla must now come back to its own partner, its route becomes a closed tour, which
lengthens the round.
For BB codes the round becomes 28--37\% longer (Table~\ref{tab:roundtime}).
For our circulant-permutation codes it becomes 9--22\% longer, since their unflagged routes already
end close to home.

\rev{Both electrons of the singlet could also walk, each visiting part of the check before the two meet
for the Bell measurement; the flag still detects a bit flip on either electron.
For the surface code this does not shorten the round.
In CAbLECAR's schedule, which measures the $X$ and $Z$ checks in one subcycle~\cite{cablecar}, every
data qubit already takes part in a gate at every step of the walk, so a second walking electron
cannot finish the gates sooner.
For qLDPC codes, whose long walks leave data qubits waiting, splitting each walk between the two
electrons could shorten the round; for simplicity we keep $b$ in the generator throughout.}

\begin{table}[t]
\caption{Syndrome-round duration with 50\,ns per shuttle hop.
CPM: the circulant-permutation codes of Sec.~\ref{sec:codes}.}\label{tab:roundtime}
\begin{ruledtabular}\begin{tabular}{lccc}
Code & CAbLECAR & Singlet flag & Change\\\hline
BB $\nkd{72}{12}{6}$ & 9.6\,$\mu$s & 12.3\,$\mu$s & +28\%\\
Gross $\nkd{144}{12}{12}$ & 11.2\,$\mu$s & 14.5\,$\mu$s & +30\%\\
BB $\nkd{288}{12}{18}$ & 14.1\,$\mu$s & 19.3\,$\mu$s & +37\%\\
CPM, $w=8$--$12$ & 11.4--17.3\,$\mu$s & 13.2--19.1\,$\mu$s & +9--22\%\\
\end{tabular}\end{ruledtabular}
\end{table}

\subsection{Noise model and decoding}\label{sec:noise}

We use CAbLECAR's device model, simulated as a Pauli noise circuit in stim~\cite{gidney2021}.
In both circuits the CZ is the only two-qubit gate.
In an $X$ check the data qubit takes a Hadamard before and after each CZ (Fig.~\ref{fig:circuit});
the two Hadamards that meet on a data qubit between consecutive $X$-check gates cancel, leaving two
per data qubit per round.
Each Hadamard is a single-qubit gate with its own noise and fits in an idle gap of the schedule, so
the round time is unchanged.
Every two-qubit gate is followed by two-qubit depolarising noise with probability
$p=10^{-3}$, and every single-qubit gate by single-qubit depolarising noise with probability $p/10$.
Initialisation prepares the wrong state, and readout reports the wrong outcome, each with probability $p$.
Each shuttle hop of $10\,\mu$m applies a phase flip with probability $p_{\rm sh}$ and a bit flip with
probability $p_x$, which we set to $0$, $p_{\rm sh}/10$ or $p_{\rm sh}/3$, and moving a qubit into or
out of a readout or interaction zone applies a phase flip with probability $p_{\rm sh}/10$.
Idle qubits dephase with $T_2=10$\,ms: a qubit idle for time $t$ receives a phase flip with
probability $(1-e^{-t/T_2})/2$.
\rev{Singlet preparation uses a simplified model.
Before the pair is formed, each of its two electrons, $a$ and $b$, takes a bit flip with probability
$p_s$ ($p_s=p$ by default), exactly as an initialised ancilla does in CAbLECAR; the pairing itself is
ideal.}
For the high-rate codes in Sec.~\ref{sec:mc} we also use a \emph{uniform} model, to test what the
higher circuit distance buys.
In the device model only the shuttle errors vary, while gate, readout and singlet errors stay at
$10^{-3}$, so lowering $p_{\rm sh}$ cannot take the circuit far below threshold.
In the uniform model every error rate of the device model is scaled together with a single $p$:
two-qubit gates, initialisation, readout, singlet preparation and flag readout fail with probability
$p$ and single-qubit gates with $p/10$, each shuttle hop applies a phase flip with probability $p$ and
a bit flip with probability $p/10$, moving into or out of a zone applies a phase flip with probability
$p/10$, and $T_2=10\,\text{ms}\times(10^{-3}/p)$.
At $p=10^{-3}$ the uniform model is the device model at $p_{\rm sh}=10^{-3}$ and $p_x=p_{\rm sh}/10$,
with every parameter equal.
Far enough below threshold the logical error rate then falls as $p^{\lceil d_{\rm circ}/2\rceil}$,
so a higher circuit distance should show up as a gain that grows polynomially as $p$ falls.
Flag outcomes are decoded jointly with the syndromes as ordinary detectors, using BP+OSD over the
full measurement history~\cite{panteleev2021,roffe2020}.
BP uses the min-sum rule with scaling factor $0.625$, a serial schedule and at most 200 iterations,
and OSD uses the combination-sweep method at order 4.
Both circuits are decoded with identical settings.

\rev{\paragraph*{Simulation methods.}
Each simulation is a memory experiment of $r=d$ rounds of syndrome extraction (6 for BB72, 12 for
the Gross code), with the data prepared and finally read out in the $X$ or the $Z$ basis.
We write the noisy circuit as a stim circuit with \texttt{DETECTOR} and \texttt{OBSERVABLE\_INCLUDE}
annotations~\cite{gidney2021}, in which each flag readout is a single-measurement detector.
Shots are drawn with stim's detector sampler, and the decoder works from the detector error model
that stim derives from the same circuit.
A shot fails if the decoder's correction gets any of the $k$ logical observables of that basis
wrong.
Each point is sampled in batches until it has about 100 failures or reaches its shot budget, and
some points are topped up with further independent batches, so each point has between
$2\times10^3$ and $2\times10^7$ shots.
From $f$ failures in $N$ shots we report the logical error rate per round,
$P_L=1-(1-f/N)^{1/r}$.
The $X$ and $Z$ bases are separate experiments.
For BB72 and the surface codes in Figs.~\ref{fig:bb}, \ref{fig:surface} and~\ref{fig:singlet}, we
combine the failure and shot counts from the $X$- and $Z$-basis experiments before applying the
per-round conversion.
The Gross-code device-model sweep in Fig.~\ref{fig:bb} uses only $X$-basis memory experiments,
whereas Figs.~\ref{fig:ldpc} and~\ref{fig:large} show the two bases separately.
Figure~\ref{fig:eff} reports the block failure rate per round normalized by the number of logical
qubits, $(P_L^X+P_L^Z)/k$, rather than an individual-logical-qubit failure probability.
Error bars are 95\% Clopper--Pearson intervals on $f/N$, mapped to a per-round rate in the same way;
where no failure is recorded we plot the 95\% upper limit.}

\section{Codes of fixed distance and increasing weight}\label{sec:codes}

To see how check weight affects circuit distance, we need codes that differ in weight but share a
distance.
Otherwise a change in circuit distance could simply reflect a change in code distance.
The known BB and circulant-permutation codes change both together, so we constructed new ones.

Circulant-permutation (CPM) pair-partition codes~\cite{okada2026,ultrahighrate2026} have check
matrices built from blocks that are cyclic shifts of the $P\times P$ identity matrix.
The number of block columns sets the check weight $w$.
We fixed three block rows, chose the shifts so that the code's Tanner graph has no 4-cycles (girth
6), and kept the smallest $P$ for which the code's distance certifies at exactly 6.
Certification was exhaustive: no logical of weight 4 or less exists, and weight-6 logicals are
exhibited explicitly.
All instances are available online (Appendix~\ref{app:codes}).

Our reference set is BB $\nkd{72}{12}{6}$ ($w=6$) and the new codes $\nkd{104}{30}{6}$ ($w=8$),
$\nkd{170}{72}{6}$ ($w=10$) and $\nkd{228}{118}{6}$ ($w=12$).
Their encoding rates $k/n$ are $0.17$, $0.29$, $0.42$ and $0.52$: higher weight buys a higher rate
at the same distance.
To check that our conclusions are not artefacts of one instance, we built two further independent
codes at each weight, plus one at a larger $P$ ($\nkd{136}{38}{6}$, $\nkd{290}{120}{6}$,
$\nkd{444}{226}{6}$).
For codes of higher distance we also use three earlier circulant-permutation codes, the BB codes
$\nkd{90}{8}{10}$ and $\nkd{108}{8}{10}$~\cite{bravyi2024}, and new codes of distance 8 and 10 at
check weights 6 to 12.
We found these by random search over the same construction, with three or four block rows and no
girth condition, certifying each code distance exactly with the solver of Ref.~\cite{cheng2026}
(Sec.~\ref{sec:distance}), and kept the smallest code at each distance and weight; for weights 8 and
10 at distance 8 we also kept a larger one.
All of the high-rate codes are listed in Table~\ref{tab:codes}.

\begin{table}[t]
\caption{The high-rate circulant-permutation codes used in this work. $w$: check weight; $J$: block
rows (column weight); $P$: circulant size.
Instances a, b, c share $\nkd{n}{k}{d}$ but have independently chosen shifts.
BB $\nkd{72}{12}{6}$ ($w=6$, $k/n=0.17$) is shown for comparison.}\label{tab:codes}
\begin{ruledtabular}\begin{tabular}{lccccc}
Code & $w$ & $J$ & $P$ & $k/n$ & Used for\\\hline
BB $\nkd{72}{12}{6}$ & 6 & -- & -- & 0.17 & both\\[2pt]
$\nkd{104}{30}{6}$ a, b, c & 8 & 3 & 13 & 0.29 & both\\
$\nkd{136}{38}{6}$ & 8 & 3 & 17 & 0.28 & distance\\
$\nkd{170}{72}{6}$ a, b, c & 10 & 3 & 17 & 0.42 & both\\
$\nkd{290}{120}{6}$ & 10 & 3 & 29 & 0.41 & distance\\
$\nkd{228}{118}{6}$ a, b, c & 12 & 3 & 19 & 0.52 & distance\\
$\nkd{444}{226}{6}$ & 12 & 3 & 37 & 0.51 & distance\\[2pt]
$\nkd{104}{30}{8}$ & 8 & 3 & 13 & 0.29 & both\\
$\nkd{176}{48}{10}$ & 8 & 3 & 22 & 0.27 & both\\
$\nkd{180}{42}{12}$ & 10 & 4 & 18 & 0.23 & distance\\[2pt]
$\nkd{66}{4}{8}$ & 6 & 3 & 11 & 0.06 & distance\\
$\nkd{136}{38}{8}$ & 8 & 3 & 17 & 0.28 & distance\\
$\nkd{230}{96}{8}$ & 10 & 3 & 23 & 0.42 & distance\\
$\nkd{290}{120}{8}$ & 10 & 3 & 29 & 0.41 & distance\\
$\nkd{132}{50}{8}$ & 12 & 4 & 11 & 0.38 & distance\\
$\nkd{444}{226}{8}$ & 12 & 3 & 37 & 0.51 & distance\\
$\nkd{66}{4}{10}$ & 6 & 3 & 11 & 0.06 & distance\\
$\nkd{204}{74}{10}$ & 12 & 4 & 17 & 0.36 & distance\\
\end{tabular}\end{ruledtabular}
\end{table}

\section{Circuit distance}\label{sec:distance}

\paragraph*{Exact computation.}
Computing Eq.~\eqref{eq:dcirc} is a minimum-weight codeword problem, which is hard in general, and
the usual heuristic searches only return upper bounds.
In our circuits stim's built-in search~\cite{gidney2021} missed the true minimum in seven of the eighteen cases where we have both: three of twelve at
distance 6 and four of six at distances 8 and 10.
Such upper bounds are often not tight: a recent audit of 26 papers found published circuit distances,
or bounds read as distances, above the certified values in ten of them, by up to eight~\cite{cheng2026}.
We instead pose Eq.~\eqref{eq:dcirc} as a weighted MaxSAT problem.
Each fault mechanism is a Boolean variable.
Each detector contributes a parity constraint, written as a chain of XOR clauses, requiring it not
to fire.
One clause requires at least one logical to flip, and each mechanism carries a unit penalty for
being used.
The RC2 solver~\cite{rc2} returns a minimum-weight solution together with a proof that nothing
smaller exists, so the result is exact.
We check each returned solution outside the solver: adding up, modulo 2, the detectors and
logical observables flipped by the chosen fault mechanisms, we confirm that no detector fires and at
least one observable flips, so the solution is a genuine undetectable logical fault.
Optimality rests on the solver's proof; as a further consistency check, every exact value is at or
below the upper bound from stim's search wherever both are available.
Our circuits have $2\times10^3$ to $3.4\times10^4$ distinct fault mechanisms, and MaxSAT solve
times range from seconds to days, growing mainly with the answer rather than the circuit size.

Every entry of Tables~\ref{tab:dist}--\ref{tab:distsurface} is certified by an independent
solver~\cite{cheng2026} that searches directly over sets of fault mechanisms of increasing size and
proves each lower bound by exhaustion, in seconds to hours.
Each minimum-weight fault set it returns is replayed in the noiseless stim circuit.
We validated it against MaxSAT on the codes of Table~\ref{tab:dist}: it reproduced all 45 values
that MaxSAT settled, and settled the seven that MaxSAT did not finish within about a week.
For codes of distance 8 to 12 (Table~\ref{tab:distlarge}) we use the solver of Ref.~\cite{cheng2026}
alone.

\begin{table}[t]
\caption{Exact circuit distance, written ``$X$-basis / $Z$-basis''.
All codes have code distance 6; $w$ is the check weight.
Instances a, b, c at weights 8--12 are independent codes of the same size.
CAbLECAR denotes the mobile-ancilla walk with the same CZ gates as the flag.
Every entry is certified by the solver of Ref.~\cite{cheng2026} (Sec.~\ref{sec:distance}).}\label{tab:dist}
\begin{ruledtabular}\begin{tabular}{llcc}
$w$ & Code & CAbLECAR & Singlet flag\\\hline
6 & BB $\nkd{72}{12}{6}$ & 5 / 5 & 6 / 6\\[2pt]
8 & $\nkd{104}{30}{6}$ a & 4 / 4 & 5 / 5\\
  & \phantom{$\nkd{104}{30}{6}$} b & 4 / 3 & 5 / 5\\
  & \phantom{$\nkd{104}{30}{6}$} c & 3 / 4 & 5 / 5\\
  & $\nkd{136}{38}{6}$ & 4 / 4 & 6 / 6\\[2pt]
10 & $\nkd{170}{72}{6}$ a & 3 / 4 & 5 / 5\\
   & \phantom{$\nkd{170}{72}{6}$} b & 4 / 4 & 5 / 5\\
   & \phantom{$\nkd{170}{72}{6}$} c & 3 / 3 & 5 / 5\\
   & $\nkd{290}{120}{6}$ & 4 / 5 & 6 / 6\\[2pt]
12 & $\nkd{228}{118}{6}$ a & 1 / 3 & 3 / 3\\
   & \phantom{$\nkd{228}{118}{6}$} b & 1 / 1 & 2 / 2\\
   & \phantom{$\nkd{228}{118}{6}$} c & 2 / 2 & 2 / 2\\
   & $\nkd{444}{226}{6}$ & 5 / 5 & 6 / 5\\
\end{tabular}\end{ruledtabular}
\end{table}

\begin{table}[t]
\caption{Exact circuit distance of the higher-distance codes, ``$X$-basis / $Z$-basis'', from the
solver of Ref.~\cite{cheng2026} with the round-collapsed lower bound; every minimum-weight fault set
is replayed in stim.
$w$: check weight; $^*$: constructed for this work (Sec.~\ref{sec:codes}).}\label{tab:distlarge}
\begin{ruledtabular}\begin{tabular}{lccc}
Code & $w$ & CAbLECAR & Singlet flag\\\hline
$\nkd{66}{4}{8}^*$ & 6 & 5 / 6 & 7 / 7\\
$\nkd{104}{30}{8}$ & 8 & 4 / 4 & 5 / 6\\
$\nkd{136}{38}{8}^*$ & 8 & 4 / 4 & 6 / 6\\
$\nkd{230}{96}{8}^*$ & 10 & 4 / 4 & 6 / 5\\
$\nkd{290}{120}{8}^*$ & 10 & 5 / 5 & 6 / 6\\
$\nkd{132}{50}{8}^*$ & 12 & 4 / 4 & 6 / 5\\
$\nkd{444}{226}{8}^*$ & 12 & 3 / 6 & 5 / 7\\[2pt]
$\nkd{66}{4}{10}^*$ & 6 & 5 / 6 & 9 / 9\\
BB $\nkd{90}{8}{10}$ & 6 & 6 / 6 & 9 / 9\\
BB $\nkd{108}{8}{10}$ & 6 & 6 / 7 & 10 / 10\\
$\nkd{176}{48}{10}$ & 8 & 5 / 6 & 7 / 7\\
$\nkd{204}{74}{10}^*$ & 12 & 6 / 6 & 8 / 8\\[2pt]
Gross $\nkd{144}{12}{12}$ & 6 & 7 / 7 & 10 / 10\\
$\nkd{180}{42}{12}$ & 10 & 8 / 8 & 10 / 10\\
\end{tabular}\end{ruledtabular}
\end{table}

\begin{table}[t]
\caption{Exact circuit distance of the rotated surface code over $d$ rounds, ``$X$-basis /
$Z$-basis'', computed as in Table~\ref{tab:distlarge}.}\label{tab:distsurface}
\begin{ruledtabular}\begin{tabular}{lcc}
Code & CAbLECAR & Singlet flag\\\hline
$d=3$ & 3 / 3 & 3 / 3\\
$d=5$ & 5 / 5 & 5 / 5\\
$d=7$ & 7 / 7 & 7 / 7\\
$d=9$ & 9 / 9 & 9 / 9\\
\end{tabular}\end{ruledtabular}
\end{table}

\begin{figure*}[t]\centering
\includegraphics[width=0.85\textwidth]{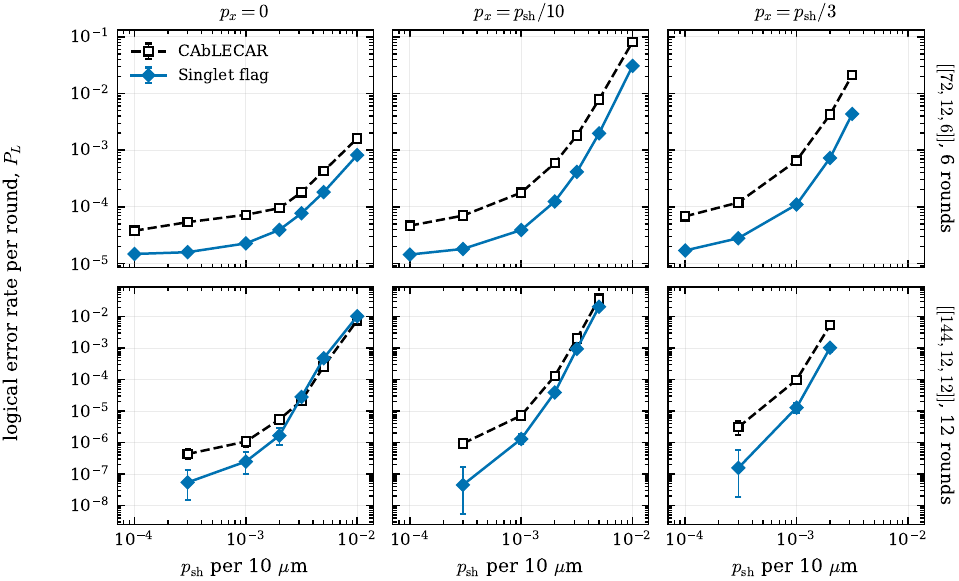}
\caption{Logical error rate per round against shuttle phase-flip probability $p_{\rm sh}$ per hop,
for BB $\nkd{72}{12}{6}$ (top) and the Gross code (bottom), with shuttle bit-flip probability
$p_x=0$, $p_{\rm sh}/10$ and $p_{\rm sh}/3$ (columns).
Error bars: 95\% confidence intervals.}\label{fig:bb}
\end{figure*}

Tables~\ref{tab:dist} and~\ref{tab:distlarge} contain the central result: the first across
check weights at fixed code distance, the second for codes of distance 8 to 12, which shows that it
carries over to larger codes.
Without the flag, the circuit distance is 5 at weight 6 and falls to 3--4\rev{, in at least one basis,}
at weights 8 and 10.
At weight 12 the three instances of $\nkd{228}{118}{6}$ fall to 1--3, and two of them have circuit
distance 1 in the $X$ basis: a single fault causes an undetected logical error.
The larger weight-12 code $\nkd{444}{226}{6}$ keeps 5 in both bases.
Apart from it and the weight-6 BB code, every unflagged circuit corrects at most one fault in at least
one basis, half of what the code promises.

With the singlet flag, the weight-6 BB code recovers its full distance of 6, and every weight-8 and
weight-10 instance reaches at least 5, which corrects the same two faults as the code.
The larger instances $\nkd{136}{38}{6}$ and $\nkd{290}{120}{6}$ reach the full 6 in both bases.
At weights 8 and 10 the flag lifts every instance, whatever the unflagged value was, which makes the
circuit distance far more predictable.
At weight 12 the flagged circuits of $\nkd{228}{118}{6}$ stay at 2--3 and the gain is erratic: two
units in the $X$ basis of instance a and none in its $Z$ basis, one in each basis of instance b, and
none in either basis of instance c.
The larger $\nkd{444}{226}{6}$ reaches 6 and 5 with the flag.

Table~\ref{tab:distlarge} gives the circuit distances of codes of distance 8 to 12.
For the three that we simulate in Sec.~\ref{sec:mc}, $\nkd{104}{30}{8}$, $\nkd{176}{48}{10}$ and the
Gross code, every circuit keeps only half to three-fifths of its code distance without the flag: 4 at distance 8,
5--6 at distance 10, and 7 for the Gross code.
With the flag, the distance-10 code reaches 7 in both bases and the Gross code 10 in both bases, three
more than CAbLECAR, so the flagged Gross circuit corrects four faults where CAbLECAR corrects three.
That is the circuit distance of the standard fixed-coupler schedule of the Gross
code~\cite{bravyi2024,cheng2026}, although the two fault models differ.
For $\nkd{180}{42}{12}$ CAbLECAR reaches 8 in both bases and the flagged circuit 10, a gain
of two.
The flag gains least, one unit, in the $X$ basis of the distance-8 code and the $Z$ basis of the
distance-10 code.
In the former the minimum logical fault is two pairs of bit flips, each pair on one ancilla and invisible to its flag, plus one data
fault.
Heuristic searches would have misled us here.
Randomised BP+OSD searches on the Gross code had found nothing below 11 for the flagged circuit,
whose true value is 10, and stim's search overestimated four of the six distance-8 and distance-10
values it was run on, by up to two.

The new codes of distance 8 and 10 show the pattern of Table~\ref{tab:dist} one step up in distance.
Without the flag, every distance-8 circuit keeps only 3 to 5 in at least one basis, so five of the
seven correct a single fault, and $\nkd{444}{226}{8}$ has circuit distance 3 in the $X$ basis.
With the flag every distance-8 circuit reaches at least 5 and corrects two faults.
At distance 10 the flag gains 3 to 4 units at weight 6 and 2 at weight 12, and the weight-6 BB code
$\nkd{108}{8}{10}$ keeps its full distance of 10 with the flag, against 6 and 7 without.

\paragraph*{Surface codes.}
For the rotated surface code CAbLECAR already keeps the full code distance, and so does the flag
(Table~\ref{tab:distsurface}): with weight-4 checks no hook lowers the circuit distance, so there is
nothing for the flag to recover.
Any gain of the flag on the surface code must therefore come from the prefactor $A$ rather than the
exponent.

\section{Logical error rates}\label{sec:mc}

\begin{figure*}[t]\centering
\includegraphics[width=0.85\textwidth]{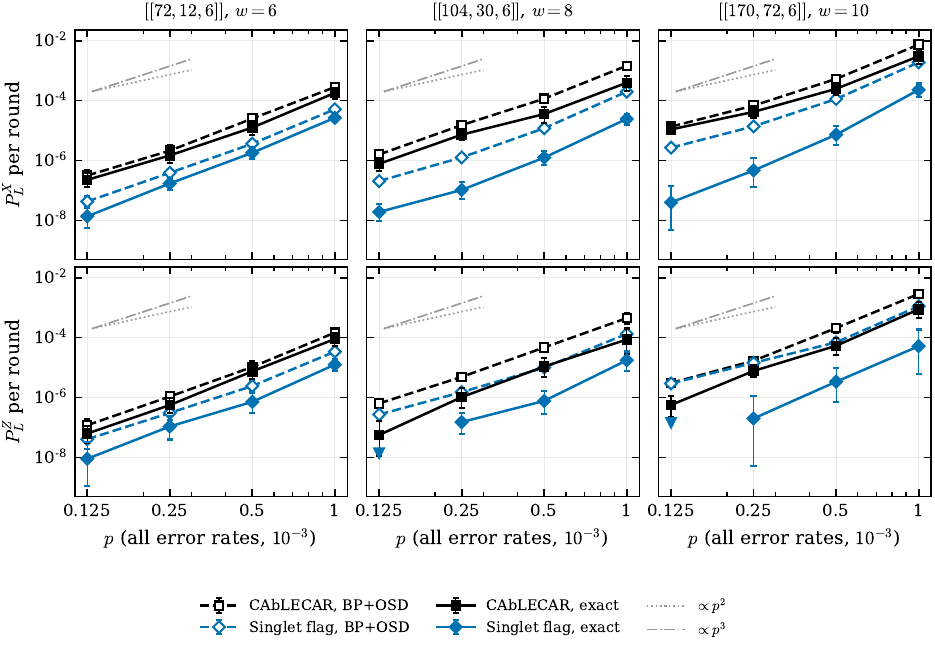}
\caption{Distance-6 codes under the uniform model, in which every error source scales with $p$ and
transport bit flips occur with probability $p/10$ per hop: BB72 ($w=6$) and the weight-8 and
weight-10 codes (instance a).
Logical error rate per round over 6 rounds in the $X$ (top) and $Z$ (bottom) basis.
Black squares: CAbLECAR; blue diamonds: singlet flag.
Open markers, dashed lines: BP+OSD; filled markers, solid lines: exact minimum-weight decoding
(Sec.~\ref{sec:decoder}).
Every shot is decoded both ways.
Triangles: 95\% upper limits where exact decoding recorded no failure.
Grey lines: the slopes $p^2$ and $p^3$.
Error bars: 95\% confidence intervals.}\label{fig:ldpc}
\end{figure*}

The circuit distances of Sec.~\ref{sec:distance} count how many faults each circuit can correct.
This section asks what that buys in the logical error rate, through three questions.
First, does the flag lower the logical error rate on a realistic device, where the noise is set by
measured hardware parameters rather than by the circuit distance?
Second, does the higher circuit distance show up as a steeper fall of the logical error rate with
$p$, with the exponent $\lceil d_{\rm circ}/2\rceil$?
Third, what happens at larger code distance, where that exponent is out of reach?
\rev{Unless otherwise stated, logical error rates are obtained with BP+OSD using identical settings
for both circuits; Fig.~\ref{fig:ldpc} also includes exact-decoding results.}

\paragraph*{A realistic device.}
We first simulate BB72 and the Gross code under the device model of Sec.~\ref{sec:noise}
(Fig.~\ref{fig:bb}).
For BB72 the flag lowers the logical error rate at every sampled point: by $2.6$--$4.8\times$ with
$p_x=p_{\rm sh}/10$, by $4.0$--$6.0\times$ with $p_x=p_{\rm sh}/3$, and still by $2.0$--$3.4\times$
when shuttling causes no bit flips at all.
For the Gross code with shuttle bit flips the gain is $1.8$--$7.7\times$, and larger at the lowest
$p_{\rm sh}$, where few failures limit the precision ($\sim\!20\times$, 95\% interval $5$--$86$, at
$p_{\rm sh}=3\times10^{-4}$).
Without shuttle bit flips the Gross code gains $3$--$8\times$ up to $p_{\rm sh}=2\times10^{-3}$ but
\emph{loses} above it, by $1.3$--$1.8\times$ at $p_{\rm sh}=3.16\times10^{-3}$--$10^{-2}$: there the
flag has no bit flips to catch, while its rounds, $30\%$ longer, collect more transport phase flips.

\begin{figure*}[t]\centering
\includegraphics[width=0.85\textwidth]{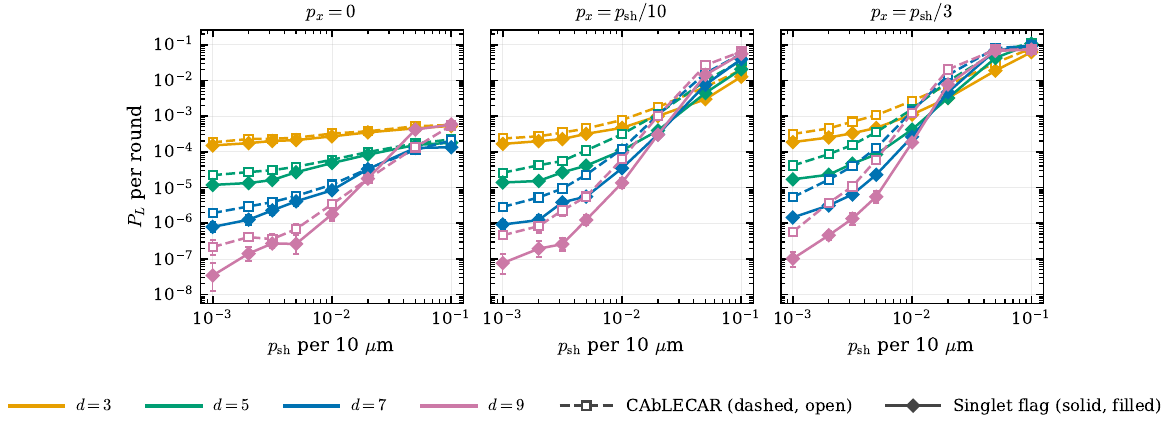}
\caption{Rotated surface code, $d=3$--$9$, over $d$ rounds, against shuttle phase-flip probability
$p_{\rm sh}$ per hop, with the $X$- and $Z$-basis memories pooled.
Dashed lines with open squares: CAbLECAR; solid lines with filled diamonds: singlet flag; colour:
code distance.
Columns: shuttle bit-flip probability $p_x=0$, $p_{\rm sh}/10$, $p_{\rm sh}/3$.
Error bars: 95\% confidence intervals.}\label{fig:surface}
\end{figure*}

The surface code is a useful reference (Fig.~\ref{fig:surface}).
Its weight-4 checks lose no circuit distance, with or without the flag
(Table~\ref{tab:distsurface}), so here the flag can lower only the prefactor $A$.
It still does, and the gain grows with distance and bit-flip fraction: with $p_x=p_{\rm sh}/3$ and
$p_{\rm sh}\le10^{-2}$ it is $1.7$--$2.4\times$ at $d=3$, $2.4$--$4.2\times$ at $d=5$,
$3.7$--$6.2\times$ at $d=7$ and up to $11\times$ at $d=9$.
Without shuttle bit flips it falls to $1.2$--$2.4\times$ for $d\le7$ and $p_{\rm sh}\le10^{-2}$, since
weight-4 hooks are already nearly harmless, and at very high transport phase-flip rates
($p_{\rm sh}=5\times10^{-2}$) the flagged $d=7$ and $d=9$ codes are worse than unflagged ones.

\rev{Under the device model, the flag improves the simulated qLDPC memories across the sampled
regimes with shuttle bit flips.
The uniform-noise comparisons below and the decoder tests of Sec.~\ref{sec:decoder} show that BP+OSD
leaves some of the available protection unrealized.
In the tested comparisons, exact decoding improves the flagged circuit more than CAbLECAR, increasing
the measured advantage of the flag.}

\begin{figure*}[t]\centering
\includegraphics[width=0.85\textwidth]{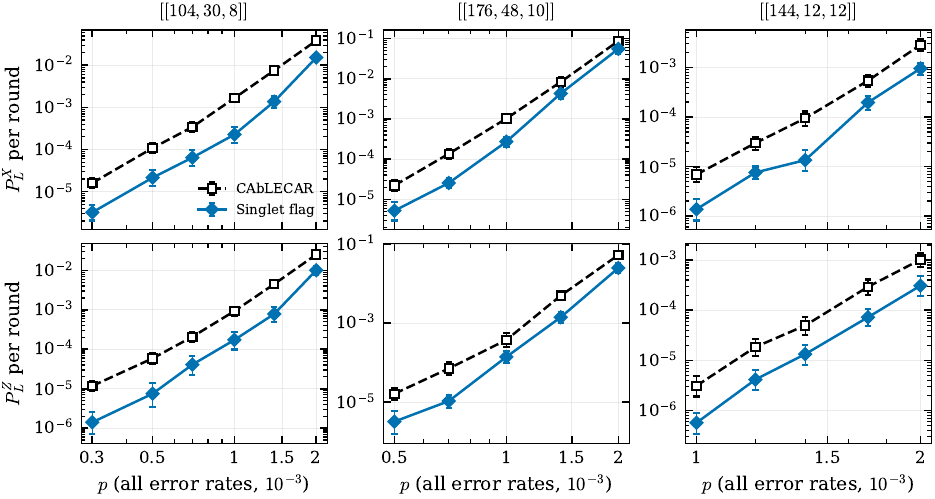}
\caption{Higher-distance codes under the uniform model: the circulant-permutation codes
$\nkd{104}{30}{8}$ and $\nkd{176}{48}{10}$ ($w=8$) and the Gross code $\nkd{144}{12}{12}$ ($w=6$).
Logical error rate per round in the $X$ (top) and $Z$ (bottom) basis.
Error bars: 95\% confidence intervals.}\label{fig:large}
\end{figure*}

\paragraph*{Looking for the exponent: distance-6 codes.}
The device model cannot answer the second question.
Its gate, readout and singlet errors are fixed at $10^{-3}$, so lowering $p_{\rm sh}$ never takes the
circuit far enough below threshold for the lowest-weight failures to dominate, which is where the
logical error rate falls as $p^{\lceil d_{\rm circ}/2\rceil}$.
We therefore use the uniform model of Sec.~\ref{sec:noise}, in which every error source scales with
a single $p$, and the distance-6 codes, which are small enough to sample down to
$p=3.1\times10^{-5}$ at weight 8 and $6.3\times10^{-5}$ at weight 10 with BP+OSD;
Fig.~\ref{fig:ldpc} shows $p=1.25\times10^{-4}$--$10^{-3}$, where we also decoded every shot exactly.
For the weight-8 code the circuit distances predict $p^2$ for CAbLECAR ($d_{\rm circ}=4$) and $p^3$
with the flag ($d_{\rm circ}=5$), in both bases, and so do the weight-10 distances.
The flag's gain should therefore grow as $1/p$, tenfold for every tenfold fall in $p$.
Seeing that growth would show that the extra circuit distance gives a polynomial reduction of the
logical error rate, not only a constant factor.

At weight 8 the flag lowers the logical error rate at every sampled $p$ below saturation: by
$8$--$24\times$ in the $X$ basis and $2.8$--$5.6\times$ in the $Z$ basis for $p\le10^{-3}$, shrinking
towards $1$ only as both circuits approach saturation near $5\times10^{-3}$.
The slopes, however, barely separate.
In the $X$ basis, fits at the three lowest error rates ($3.1\times10^{-5}$--$1.25\times10^{-4}$, each
point weighted by its number of failures) give $2.08\pm0.16$ for CAbLECAR, on its predicted exponent
of 2, but only $2.49\pm0.16$ for the flag, so the gain grows only weakly as $p$ falls.
This is a limitation of BP+OSD, not of the circuit: at $p=3.1\times10^{-5}$ about 40\% of the flag's
logical error rate comes from fault pairs that the flagged circuit can correct but BP+OSD decodes
wrongly, which scale as $p^2$; exact decoding corrects all of them (Sec.~\ref{sec:decoder}).
In a separate run at $p=1.25\times10^{-4}$--$10^{-3}$ that decodes the same shots both ways
(Fig.~\ref{fig:ldpc}), exact decoding raises the weight-8 gain from $7.3$--$12\times$ to
$16$--$70\times$ in the $X$ basis and from $2.4$--$4.7\times$ to $4.9$--$14\times$ in the $Z$ basis.
In the $X$ basis the exact-decoding gain also grows as $p$ falls, from $16\times$ at $p=10^{-3}$ to
$41$--$70\times$ at $p\le2.5\times10^{-4}$, whereas the BP+OSD gain on the same shots does not.

At weight 10, BP+OSD hides the flag's advantage in the $Z$ basis.
With BP+OSD the flag wins by $1.7$--$2.9\times$ at $p=5\times10^{-4}$--$2\times10^{-3}$, but below
$p\approx3\times10^{-4}$ the two circuits tie (ratios $0.81$--$1.25$, 95\% intervals all including 1),
although the flag has the larger $Z$ circuit distance ($5$ against $4$, Table~\ref{tab:dist}).
Decoding the same shots exactly removes the tie: the flag wins by $16$--$39\times$ at
$p=2.5\times10^{-4}$--$10^{-3}$ and records no failure in $4.3\times10^6$ shots at $1.25\times10^{-4}$,
against seven for CAbLECAR (Fig.~\ref{fig:ldpc}).
BP+OSD fails 20 to 70 times as often as exact decoding on the flagged circuit, but only 2 to 5 times as
often on CAbLECAR.
Switching off one noise channel at a time at $p=2.5\times10^{-4}$ shows which faults it mishandles.
Without transport bit flips the flag's BP+OSD failure rate falls $4.5\times$ and CAbLECAR's $2.3\times$,
and the flag wins by $2.2\times$ (95\% interval $1.6$--$2.9$); without idle dephasing or transport phase
flips the tie remains.
Every single bit flip is flagged and corrected, so the failures BP+OSD makes here are combinations of
two or more faults that include a transport bit flip.

\begin{figure*}[t]\centering
\includegraphics[width=0.85\textwidth]{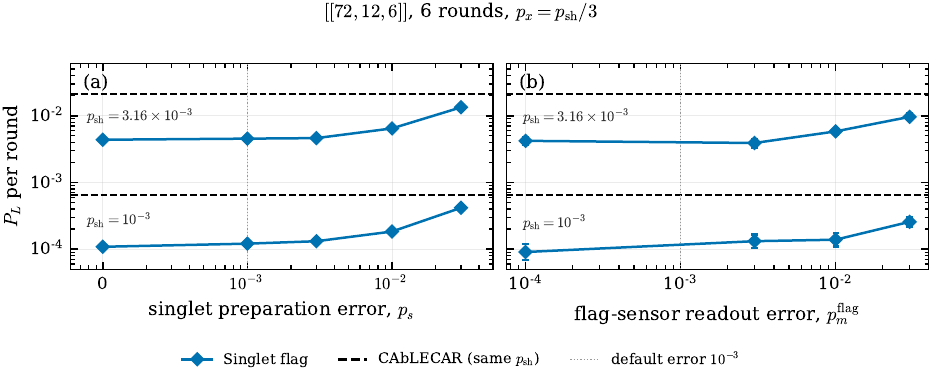}
\caption{BB $\nkd{72}{12}{6}$ with $p_x=p_{\rm sh}/3$: singlet-flag logical error rate per round
against (a) singlet-preparation error and (b) flag-sensor readout error.
Upper pair of curves: $p_{\rm sh}=3.16\times10^{-3}$; lower pair: $p_{\rm sh}=10^{-3}$.
Dashed lines: CAbLECAR at the same shuttle error.
Dotted line: the default error of $10^{-3}$.}\label{fig:singlet}
\end{figure*}

\paragraph*{Larger distance.}
For the codes of distance 8 to 12 (Table~\ref{tab:distlarge}) the asymptotic exponent is out of
reach.
With circuit distances of 4 to 10 it is 2 to 5, and the logical error rates at which those terms
would dominate need far more shots than we can afford, so we sample only
$p=3\times10^{-4}$--$2\times10^{-3}$ for $\nkd{104}{30}{8}$, $5\times10^{-4}$--$2\times10^{-3}$ for
$\nkd{176}{48}{10}$ and $10^{-3}$--$2\times10^{-3}$ for the Gross code (Fig.~\ref{fig:large}).
In this range the flag lowers the logical error rate at every point and in both bases, by
$2.5$--$8.4\times$ for $\nkd{104}{30}{8}$, $1.6$--$6.7\times$ for $\nkd{176}{48}{10}$ and
$2.8$--$7.0\times$ for the Gross code, and every 95\% interval excludes 1.
For the distance-8 and distance-10 codes the gain also grows as $p$ falls, from
$1.6$--$2.5\times$ at $p=2\times10^{-3}$ to $4$--$8\times$ at the lowest $p$ simulated.
This close to threshold, failures involve many faults, so the slopes are not the asymptotic
exponents and we do not extrapolate the trend: the distance-6 codes show that the gain can level off
at lower $p$.
For the Gross code the sampled range spans only a factor of two in $p$ and the trend is not
resolved.

\paragraph*{Weight 12.}
We do not report logical error rates for the weight-12 codes, because there BP+OSD fails even on
single faults that the circuit can correct (Sec.~\ref{sec:decoder}); for weight 12 we report only
the decoder-independent circuit distance (Table~\ref{tab:dist}).

\section{Robustness to singlet and flag errors}\label{sec:robust}

The flag adds two new ways to fail: the singlet may be prepared wrongly, and the flag may be
misread.
\rev{We vary the singlet-preparation error $p_s$ of the model of Sec.~\ref{sec:noise}, a bit flip on each
electron before pairing, and the flag-sensor readout error $p_m^{\rm flag}$, with every other error
at its default.}
Neither undoes the benefit (Fig.~\ref{fig:singlet}).
For BB72 at $p_{\rm sh}=10^{-3}$, CAbLECAR's logical error rate is $6.5\times10^{-4}$ per round.
With the flag it is $1.1\times10^{-4}$ for perfect singlets, $1.2\times10^{-4}$ at the default
singlet error of $10^{-3}$, $1.8\times10^{-4}$ at $p_s=1\%$, and still $4.2\times10^{-4}$ at
$p_s=3\%$, thirty times the gate error\rev{, at which about 6\% of pairs are wrong.
These robustness results apply to the preparation-error channel of Sec.~\ref{sec:noise}.
More general loading and separation error channels, including components not detected by the flag,
are not included.}
Flag-readout errors behave similarly: at $p_m^{\rm flag}=3\%$ the flag's logical error rate is
$2.6\times10^{-4}$, still $2.5\times$ below CAbLECAR.
At $p_{\rm sh}=3.16\times10^{-3}$ the picture is the same, with the flag below CAbLECAR's
$2.1\times10^{-2}$ by at least $1.6\times$ throughout.
The flag is useful because it catches likely faults, so occasional false alarms cost little.

\section{BP+OSD does not reach the full circuit distance}\label{sec:decoder}

A circuit's distance is only useful if the decoder exploits it.
Because we know the circuit distances exactly, we can tell decoder failures from circuit failures,
and we find that BP+OSD does not reach the full circuit distance of the flagged circuits.
\rev{In a circuit of distance $d_{\rm circ}$ every set of at most $\lfloor(d_{\rm circ}-1)/2\rfloor$ faults is
correctable: a decoder that minimises the number of faults corrects it.
A decoder that weighs faults by their probabilities, as BP+OSD does, need not, because with
nonuniform fault probabilities a likelier set of more faults can share the syndrome; whether such a
failure belongs to the decoder or to the priors we settle by exact decoding below.}
We tested this directly: we gave the production decoder the syndrome of every single fault and of
every pair of faults that share a detector, and counted the corrections that flip a logical
observable.

\paragraph*{Single faults.}
At weights 8 and 10 the decoder corrects every single fault, in both circuits and both bases.
At weight 12 it does not.
For instance a at $p=6.25\times10^{-5}$ it gets six single faults of the flagged circuit wrong in
the $X$ basis and two in the $Z$ basis, although its circuit distance of 3 guarantees that every
single fault is correctable.

\paragraph*{Fault pairs.}
The flagged weight-8 circuit has circuit distance 5, so every pair of faults is correctable.
In the $X$ basis the decoder fails on 22 of its $1.2\times10^6$ overlapping pairs.
These are not near-ties.
For every one of them the decoder returns a correction of 4 to 9 faults that is less likely, under
the decoder's own priors, than the two faults that actually occurred: BP+OSD did not find the
better explanation.
Neither raising the OSD order from 4 to 10 nor replacing OSD by localised statistics decoding
(BP+LSD~\cite{lsd}, same BP settings) corrects any of the 22.
At $p=3.1\times10^{-5}$ they contribute $2.2\times10^{-8}$ per shot, about 40\% of the flag's
observed logical error rate, and because they scale as $p^2$ they hold the flag's slope near 2
although a circuit distance of 5 allows 3 (Sec.~\ref{sec:mc}).
CAbLECAR, with circuit distance 4, fails on 246 of its $2.8\times10^5$ overlapping pairs.
\rev{For 101 of these fault pairs, the likelihood objective favours a logically incorrect
explanation, so exact optimization of the same objective does not correct them; the other 145 are
search failures of the same kind as those in the flagged circuit.}

\begin{figure*}[t]\centering
\includegraphics[width=0.85\textwidth]{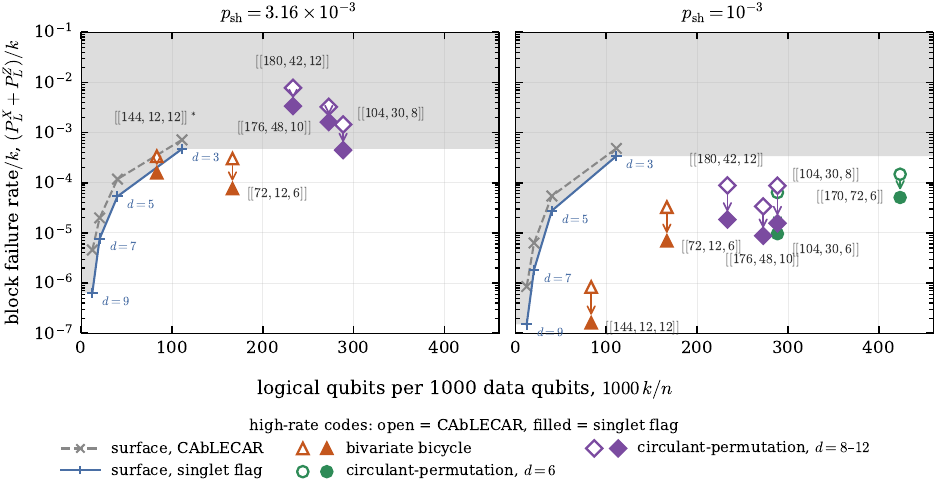}
\caption{Block failure rate per round normalised by the number of logical qubits, $(P_L^X+P_L^Z)/k$,
against logical qubits per 1000 data qubits, $1000\,k/n$, under the device model with
$p_x=p_{\rm sh}/10$, two-qubit gate and readout errors of $10^{-3}$ and single-qubit gate errors of
$10^{-4}$, in the style of Ref.~\cite{cablecar}.
$P_L^X$ ($P_L^Z$) is the probability per round that at least one of the $k$ logical qubits fails in
an $X$-basis ($Z$-basis) memory experiment; a failing shot can flip several logical qubits, so the
vertical axis is not the failure probability of an individual logical qubit.
Lines: rotated surface codes, labelled by distance, dashed for CAbLECAR and solid for the singlet flag.
Markers: high-rate codes, with shape and colour giving the code family, open for CAbLECAR and filled
for the singlet flag, joined by arrows.
Shaded: a higher logical error rate than the flagged surface code at the same encoding rate, or, beyond
$d=3$, than the flagged $d=3$ code.
The distance-6 codes appear only at $p_{\rm sh}=10^{-3}$, where our uniform model at $p=10^{-3}$ is the
same noise model with the same parameters (Sec.~\ref{sec:noise}); the Gross code at that point also
comes from the uniform runs.
$^\ast$Estimate: for the Gross code at $p_{\rm sh}=3.16\times10^{-3}$ only the $X$ basis was simulated,
and it is counted twice.
Under the uniform model at $p=10^{-3}$ and $2\times10^{-3}$ its $X$ basis fails $2.2$--$3.1$ times as
often as its $Z$ basis in both circuits, so this point overstates $(P_L^X+P_L^Z)/k$ by about
40--50\% but changes the flag's gain by less than 5\%.}\label{fig:eff}
\end{figure*}

\paragraph*{Exact decoding closes the gap.}
To confirm that these failures belong to the decoder, we decoded the same syndromes exactly, returning
the most likely set of faults consistent with each syndrome.
We first check every explanation by one or two faults, which is provably optimal whenever the
cheapest of them costs no more than any three faults could, and otherwise solve the minimum-weight
problem over the full detector error model as an integer program with HiGHS~\cite{highs}.
\rev{The exact decoder returns a most likely fault set consistent with each syndrome under the
assumed noise model.
Any remaining failure is therefore not an optimization failure for this objective.
With nonuniform fault probabilities this does not by itself guarantee that every set of at most
$\lfloor(d_{\rm circ}-1)/2\rfloor$ faults is corrected, so we test the single faults and pairs directly.}
\rev{Exact decoding corrects every single fault of the flagged weight-12 circuit in both bases, all
22 failing pairs of the flagged weight-8 circuit, and 145 of CAbLECAR's 246 failing pairs.
Thus, all of the tested low-weight BP+OSD failures in the flagged circuits are removed by exact
optimization, whereas the remaining 101 CAbLECAR failures persist under the chosen
most-likely-fault-set objective.}
\paragraph*{Exact decoding of sampled shots.}
The fault-set tests cover only one or two faults.
To see what the decoder costs in the logical error rate itself, we decoded $4.0\times10^8$ sampled
shots both exactly and with BP+OSD: BB72 and the weight-8 and weight-10 codes (instance a), under the
uniform model at $p=1.25\times10^{-4}$--$10^{-3}$, in both bases (Fig.~\ref{fig:ldpc}).
About 40\% of the shots need the integer program.
Exact decoding helps both circuits, but the flag far more.
BP+OSD fails $1.2$--$5.4$ times as often as exact decoding on CAbLECAR, but $2$--$4.5$ times as often
on the flagged BB72 circuit, $8$--$13$ times at weight 8 and $8$--$72$ times at weight 10.
The flag's gain therefore rises from $3$--$7\times$ to $5$--$17\times$ for BB72, from $2$--$12\times$
to $5$--$70\times$ at weight 8, and from $1$--$5\times$ to $13$--$90\times$ at weight 10 for
$p\ge2.5\times10^{-4}$.
In the $X$ basis the exact-decoding gain also grows as $p$ falls: from $6.8\times$ at $p=10^{-3}$ to
$17\times$ at $1.25\times10^{-4}$ for BB72, from $16\times$ to $41$--$70\times$ at weight 8, and from
$13\times$ to about $270\times$ at weight 10 (from two flag failures), whereas the BP+OSD gain on the
same shots stays within $5.5$--$7.3\times$, $7.3$--$12\times$ and $3.9$--$5.0\times$.
In the $Z$ basis the flag fails too rarely under exact decoding below $p=10^{-3}$ (none to eight
failures per point) to resolve such a trend.

\paragraph*{Consequences.}
Others have found the same limitation.
For bivariate-bicycle codes under code-capacity noise, BP+OSD fails on errors well below half the
code distance, because symmetries of the decoding graph stop BP from converging~\cite{fregona2026};
for flagged colour-code syndrome-extraction circuits under circuit-level noise, the slope of its
logical error rate at low error rates shows that it does not reach the full distance of the
code~\cite{benito2025}; and improved BP variants outperform BP+OSD at high OSD order under circuit
noise~\cite{muller2025relay}.
What the exact circuit distances add here is a clean attribution: for the flagged circuits the
shortfall is the decoder's alone, and it grows with check weight, from a few correctable pairs at
weight 8 to single faults at weight 12.
The flag makes decoding harder, since its circuits have 2.5 to 3 times as many fault mechanisms and
the decoder must read the flag outcomes correctly, so the flag gains most from a better decoder.
\rev{The baseline comparisons in Sec.~\ref{sec:mc} use identical BP+OSD settings for both circuits,
while Fig.~\ref{fig:ldpc} also reports exact-decoding results.
In those tested comparisons, more accurate decoding increases the measured advantage of the flag.}
Exact decoding shows how large the margin is: it multiplies the flag's gain over CAbLECAR by
$1$--$2.4$ for BB72, $1.5$--$6$ at weight 8 and $3$--$50$ at weight 10.

\section{Discussion}\label{sec:discussion}

The codes in Table~\ref{tab:dist} trade check weight for rate at fixed distance.
Without the flag, a shuttling device cannot cash in that rate: the higher-weight codes keep their
nominal distance on paper, but apart from the weight-6 BB code and $\nkd{444}{226}{6}$ every circuit
corrects only one fault in at least one basis, and \rev{the three instances of $\nkd{228}{118}{6}$}
not even that.
With the singlet flag, codes up to weight 10, at rate $0.42$, keep the full two-fault correction.
That is $2.5\times$ fewer physical qubits per logical qubit than the weight-6 BB code at the same
distance. When transport flips spins, it also comes with logical error rates several times lower at
fixed physical error rates.
\rev{At larger distance the flag raises the circuit distance by one to four, the Gross code's from 7
to 10 (Table~\ref{tab:distlarge}).}
When it rarely does, the weight-8 code still gains in both bases; at weight 10 BP+OSD finds only a
break-even in the $Z$ basis at low $p$, but exact decoding shows a gain of $16$--$39\times$ there too
(Fig.~\ref{fig:ldpc}).

Figure~\ref{fig:eff} compares the codes as CAbLECAR does~\cite{cablecar}: block failure rate
normalised by the number of logical qubits against encoding rate, with surface codes as the reference.
At $p_{\rm sh}=10^{-3}$ every high-rate code lies below the flagged surface-code line, so it either
encodes more logical qubits at the same logical error rate or reaches a lower logical error rate at the
same encoding rate.
The flagged Gross code reaches $1.6\times10^{-7}$ per round at 83 logical qubits per 1000 data qubits,
about the same as the flagged $d=9$ surface code ($1.5\times10^{-7}$), which encodes 12: seven times
the encoding rate at the same normalised failure rate.
The flag lowers the normalised failure rate of every high-rate code by $3$--$7\times$ here, and of the
$d=3$--$9$ surface codes by $1.4$--$5.7\times$.
\rev{The horizontal axis of Fig.~\ref{fig:eff} counts only data qubits, but syndrome extraction needs
other electrons too, and the flag needs more of them than CAbLECAR does.
CAbLECAR uses one ancilla per check and one reference spin for each readout zone in which it reads
ancillas.
With the data included, that is $3.0$--$3.2$ electrons per data qubit for the surface and BB codes
and $2.4$--$2.7$ for the circulant-permutation codes.
The flag adds a static partner for every check.
It reads ancillas at fewer nodes, but each of those nodes has two sensors, one for the ancilla and
one in the generator, and each sensor has its own reference spin.
The totals become $4.3$--$5.0$ electrons per data qubit for the surface and BB codes and
$3.0$--$3.6$ for the circulant-permutation codes, $1.3$--$1.6$ times CAbLECAR's count; all the extra
electrons are static, and each node also needs a generator.
Because the flag adds electrons to every code in a similar proportion, the comparison between
flagged codes changes little when counted in electrons rather than data qubits.
The flagged Gross code uses $4.5$ electrons per data qubit and the flagged $d=9$ surface code $5.0$,
so the Gross code's sevenfold advantage in logical qubits per data qubit becomes about $7.5$-fold
per electron, and the flagged circulant-permutation codes use $1.4$--$1.7$ times fewer electrons per
data qubit than the flagged surface codes.}
At $p_{\rm sh}=3.16\times10^{-3}$ the margin closes.
The flagged BB codes still lie at or below the surface-code line, but the circulant-permutation codes of
distance 8 to 12 lie on or above it even with the flag.

The home node of each ancilla is chosen by a simple heuristic, so our round times are upper bounds.
\rev{Unless otherwise stated, the reported logical error rates use BP+OSD.
In the tested distance-6 comparisons, exact decoding lowers the flagged logical error rates by
factors of $2$--$72$, compared with $1.2$--$5.4$ for CAbLECAR (Fig.~\ref{fig:ldpc} and
Sec.~\ref{sec:decoder}).}
All circuit distances are exact.
They hold for our schedules and fault model; a different routing, of either circuit, changes them.

\paragraph*{Outlook.}
Two extensions follow naturally.
Multi-partner flags, in which the ancilla carries several entangled partners and drops them off
along the walk, would cut each walk into short protected segments and should extend the recovery to
weight 12.
Flag-aware routing, which orders the visits so that the remaining double-fault hooks are harmless,
could close the last unit of distance on the smaller weight-8 and weight-10 codes and lift the
$X$ basis of $\nkd{104}{30}{8}$, which the flag currently leaves at 5.

\begin{acknowledgments}
This research was undertaken thanks in part to funding from the Canada First Research
Excellence Fund (Transformative Quantum Technologies) and Canada's Natural Sciences and
Engineering Research Council (NSERC).
\end{acknowledgments}

\appendix

\section{Code instances}\label{app:codes}
All circulant-permutation codes in Table~\ref{tab:codes} are available on GitHub at \url{https://github.com/KNFLWNF/singlet-flag},
each as its $J\times w$ array of circulant shifts together with $P$.
The repository also contains the scripts that build the parity-check matrices from these arrays,
certify the code distances, generate the syndrome-extraction circuits with and without the flag,
and compute the circuit distances in Table~\ref{tab:dist}, including the minimum-weight fault
sets returned by the solver.

\end{document}